\documentstyle[preprint,psfig,prb,aps]{revtex}
\begin{document}
\draft
\title{Density functional theory of the phase diagram of maximum density
droplets in two-dimensional quantum dots in a magnetic field}
\author{M.\ Ferconi}
\address{Department of Physics and Astronomy, Vanderbilt University, 
Nashville, Tennessee 37235}
\author{G.\ Vignale} 
\address{Department of Physics and Astronomy, University of Missouri-Columbia,
Columbia, Missouri 65211}

\date{\today}

\maketitle

\begin{abstract}
We present a density-functional theory (DFT) approach to the study of
the phase diagram of the maximum density droplet (MDD) in two-dimensional
quantum dots in a magnetic field. Within the lowest Landau level (LLL)
approximation, analytical expressions are derived for the values of the
parameters $N$ (number of electrons) and $B$ (magnetic field) at which the
transition from the MDD to a ``reconstructed'' phase takes place.   
The results
are then compared with  those of full  Kohn-Sham calculations, giving thus
information about  both  correlation and Landau level mixing effects. 
Our results are also contrasted with those of Hartree-Fock (HF) calculations, 
showing  that DFT  predicts a more compact  reconstructed edge,  
which is closer to the result of exact diagonalizations in the LLL.
\end{abstract} 


\narrowtext
Two-dimensional quantum dot systems, at high magnetic fields, have been
recently studied by various authors.~\cite{chakraborty92,ashoori96}
The theoretical interest in these systems
arises largely from the fact that they provide a few-electron realization of 
physical states which,  in the macroscopic limit, are responsible for the
occurrence of the quantum Hall effect.~\cite{prange90,chakraborty95}
The simplest example of such a
state is the so-called maximum density droplet (MDD) ---  which,  in the limit
of high magnetic field,  can be written as a Slater determinant of 
lowest Landau
level orbitals with angular momenta $0,1...N-1$, where $N$ is the number of
electrons.~\cite{macdonald93}  In the limit of $N \to \infty$  this coincides 
with the incompressible state of the quantum Hall effect at filling factor $\nu
=1$.  Because, within the LLL, the MDD is the {\it only} $N$-electron state of 
angular momentum $N(N-1)/2$, (and there is none with lower angular momentum)  
it follows that it must be an exact
eigenstate of the Hamiltonian

\begin{eqnarray} 
H &= & \sum_{i=1}^{N} \left[ \frac{1}{2 m^{\ast}}
\left(  {\bf p}_{i} + \frac{e}{c}  {\bf A}_{i} \right)^{2} + \frac{1}{2}
 m^{\ast} \omega_{0}^{2} r^{2}  \right]  \nonumber \\ 
&+& \frac{e^{2}}{2 k} \sum_{i=1}^{N} \sum_{j=1,j\neq i }^{N}
\frac{1}{| {\bf r}_{i} - {\bf r}_{j} | } 
 + g^{\ast} \mu_{B} B \sum_{i=1}^{N} \sigma_{i}    ,
\label{effmassH}
\end{eqnarray}
if the small Coulomb coupling
between different Landau levels is neglected. 
Here $\omega_{0}$ is the frequency of the external parabolic 
potential, ${\bf A}_{i}$ is the external vector potential, $k$ the
dielectric constant, $m^{\ast}$ the electron effective mass, $\mu_{B}$
the Bohr magneton, $g^{\ast}$ the effective $g$-factor for the Zeeman
splitting, and $\sigma_{i}$ the spin component along the
axis perpendicular to the plane of the electrons.
The question is  whether this exact eigenstate  (or rather its continuation
to finite magnetic field)   can actually be the {\it ground-state} of the 
quantum dot, in some range of magnetic fields.   
The basic physics is simple:  if the magnetic field is too large, the MDD
cannot be the ground state, because the compact arrangement of the electrons
costs too much electrostatic energy: the electrostatic stress is released
through a rearrangement of the electrons leading to a state of higher angular
momentum.  If, on the other hand,  the magnetic field is too
weak, the confinement energy
will cause the external electrons in the MDD to be transferred to the center of
the quantum dot, even though, in so doing, a higher Landau level becomes
populated at the center of the dot.  The conclusion of these arguments is that
there will exist, at most,  a ``window''  of magnetic fields in which 
the MDD is
stable.  The ``window''  shrinks with increasing electron number $N$ and it
closes up completely at a critical value $N_{c}$ of $N$ of the order of $100$.
Note that this is not in contradiction with the existence of incompressible
phases  in the macroscopic limit:  it is only telling us that such  phases will
have compressible edges.

The problem of determining quantitatively the region of stability of the MDD
has  been studied both theoretically and 
experimentally.~\cite{macdonald93,chamon94,ahn95,klein95,klein96} There exists
a disagreement between the experimental results and theoretical predictions
regarding the window of
values of magnetic field for which the MDD is the ground state.~\cite{klein95}
Correlation effects have been indicated as a possible cause for this
disagreement since they were not accounted for in the theoretical  analysis,
which was based on the Hartree-Fock approximation. The importance 
of correlations has been demonstrated for the case of small quantum
dots.~\cite{ahn95}

 Here we present an analytic treatment based on density
functional theory, which includes both exchange and correlation effects.
We shall calculate the
values of the magnetic field at which the transition  from the MDD to a new
phase takes place, as well as the angular momentum of the new phase, supposed to
lie entirely within the lowest Landau level. Within this treatment the
transition from one state to the other can be entirely described by means of
a single dimensionless parameter  
$\alpha = m^{\ast} \omega_{0}^{2} k \lambda^{3}/ e^{2} $ 
giving the strength of the  parabolic potential in terms of 
ratio between the confining energy and
the electrostatic energy existing at the typical length scale of a
magnetic length  $\lambda = \sqrt{\hbar c / (e B)}.$
We also perform a numerical evaluation, based on the solution of the
Kohn-Sham equation,
where {\it all} Landau levels are included, showing that  their inclusion
shrinks the magnetic field window of stability of the MDD. As a
byproduct of our approach we determine the maximum number of 
electrons for which the MDD can be the ground state. 

Density functional theory has already been applied successfully to
systems in the presence of a magnetic 
field,~\cite{ferconi94,stoof95,heinonen95,ferconi95} thus  establishing it
 as a useful tool for studying such systems.
The total energy of a quantum dot is 
\begin{equation}
E[n] = \int \!\! d{\bf r} \; V({\bf r}) \; n({\bf r}) + 
\frac{e^{2}}{2k} \int \!\! d{\bf r} \; 
\frac{n({\bf r}) \; n({\bf r}')}{| {\bf r - r}'|} + E_{xc}[n].
\label{etotLLL}
\end{equation}
In Eq.~(\ref{etotLLL}) we have assumed the system is within the lowest
Landau level, thus omitting the constant kinetic energy 
term $\hbar \omega_{c}/2,$ where $\omega_{c} = eB/(m^{\ast}c)$ is the
cyclotron frequency.  $V({\bf r})$ represents the parabolic confining
potential 
\begin{equation}
V({\bf r}) = \frac{1}{2} m^{\ast} \omega_{0}^{2} r^{2} = \frac{\alpha}{2} 
r^{2} \frac{e^{2}}{k \lambda^{3}},
\label{vr}
\end{equation}
The dimensionless parameter $\alpha$ gives the strength of the parabolic
potential. $E_{xc}[n]$ is the exchange-correlation energy functional. Let
now $n_{MDD}({\bf r})$ be the density of the MDD, and $n_{r}({\bf r})$
the density of the reconstructed edge obtained immediately after the
transition from the MDD takes place for example by increase of the
magnetic field. 
In the lowest Landau level picture the MDD is obtained by filling the
orbitals with angular momentum from $l = 0, $ to $l = N - 1,$ 
the system being fully spin polarized.
The reconstructed edge can be viewed as being generated  from the the
MDD by removing one electron from an orbital $\phi_{h}$ with angular
momentum $0 \leq h \leq N - 1,$ and putting it in the single-particle
orbital $\phi_{N}$ with angular momentum $N.$ Therefore 
\begin{equation}
n_{r}({\bf r}) = n_{MDD}({\bf r}) - | \phi_{h}({\bf r}) |^{2} + 
| \phi_{N}({\bf r}) |^{2}.
\label{nrnmdd}
\end{equation}
The transition from one state to the other for a given number of
electrons is {\em totally} determined by the value of the 
parameter $\alpha.$ Its critical value $\alpha_{1}$ is obtained by
solving 
\begin{equation}
E[n_{MDD}, \alpha_{1}, N] = E[n_{r}, \alpha_{1}, N].
\label{boundcond}
\end{equation}
Using Eq.~(\ref{etotLLL}) and (\ref{nrnmdd}), the energy difference
between the two states can be written as
\begin{eqnarray}
E[n_{MDD},\alpha,N] - E[n_{r},\alpha,N] &=& \nonumber \\ 
 \int \!\! d{\bf r} \; V({\bf r}) \; \left[ |\phi_{h}({\bf r})|^{2} - 
|\phi_{N}({\bf r})|^{2} \right] + \Delta E_{H} + \Delta E_{xc}. &&
\label{deltaenmddnr}
\end{eqnarray}
We are now going to evaluate separately each term on the right hand side
of Eq.~(\ref{deltaenmddnr}). The first term is easily obtained from the
second moment of the Landau orbitals
\begin{equation}
R_{l}^{2}= \int \!\! d{\bf r} \; r^{2} |\phi_{l}({\bf r})|^{2} = 
2 \lambda^{2} (|l| + 1).
\label{lan2mom}
\end{equation}
The change in Hartree energy $\Delta E_{H}$ is given by
\widetext
\begin{eqnarray}
\Delta E_{H} &=& \frac{e^{2}}{2 k} \int \!\! 
d{\bf r} \; \frac{1}{| {\bf r - r}'|} \; \left\{
2  n_{MDD}({\bf r}) \left[  |\phi_{h}({\bf r}')|^{2} - 
|\phi_{N}({\bf r}')|^{2} \right] \right. \nonumber \\
&-& \left. |\phi_{h}({\bf r})|^{2} 
|\phi_{h}({\bf r}')|^{2} - |\phi_{N}({\bf r})|^{2} 
|\phi_{N}({\bf r}')|^{2} + 2 |\phi_{h}({\bf r})|^{2} 
|\phi_{N}({\bf r}')|^{2} \right\}. 
\label{hartreeenergy}
\end{eqnarray}

\begin{figure}
\centerline{\psfig{figure=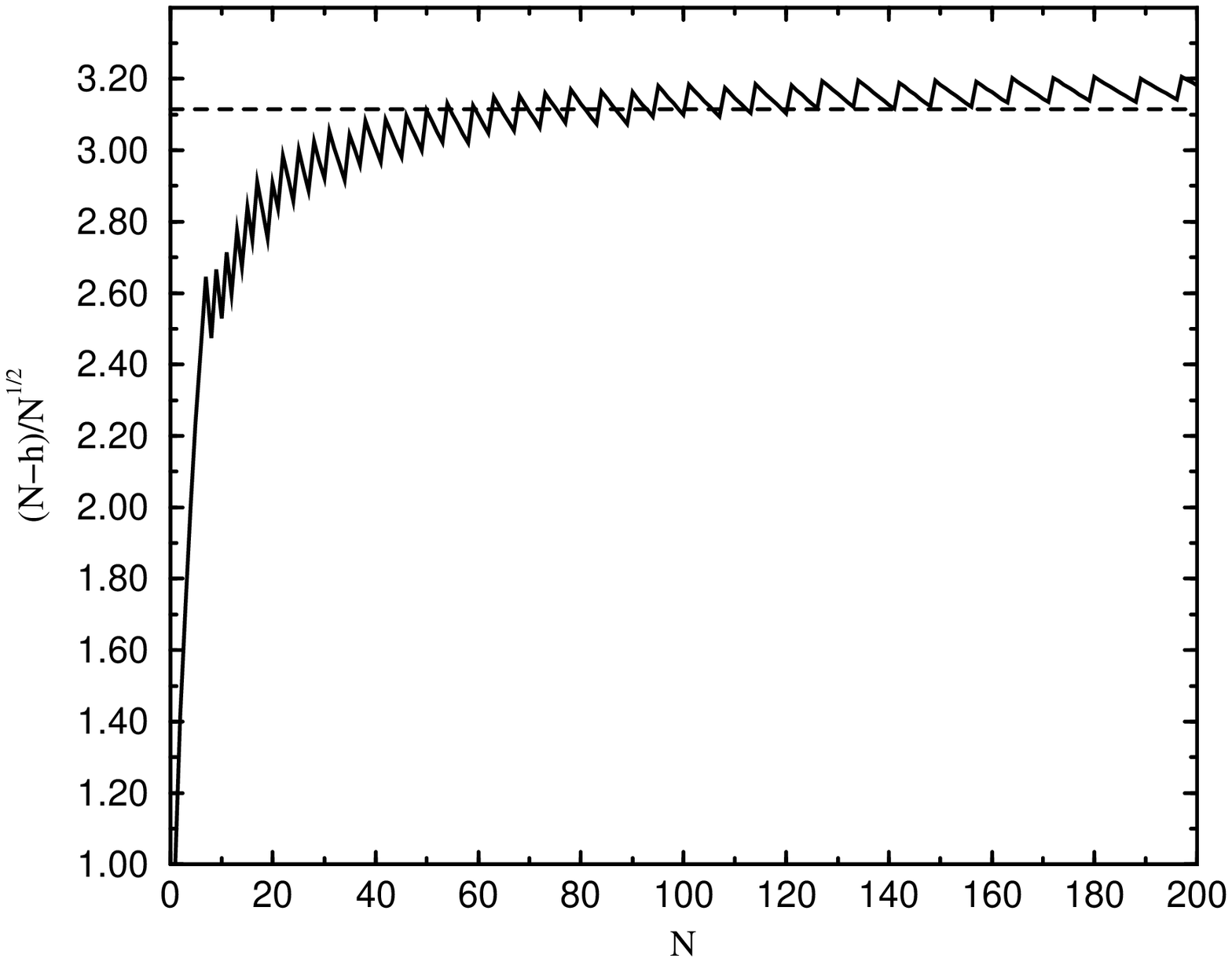,width=4.0in}}
\caption{{\it Change of angular momentum $N-h$ in going from the MDD to the
reconstructed edge scaled by $\protect\sqrt{N},$ as obtained by
maximizing  $\alpha_{1}(N; h)$ with respect to $h.$
The dashed line represents the 
asymptotic value  of Eq.~(\protect\ref{nminh}).}}
\label{holepredfig}
\end{figure}

\narrowtext
In order to proceed we approximate the MDD as a system with uniform 
density, having the $N$ electrons in a disk of radius 
$R_{MDD} = \sqrt{2 N} \lambda. $
The electrostatic potential generated by a disk of radius $R$
of uniform charge $n_{0}$ is
\begin{eqnarray}
V_{H}(r)&=&  n_{0} \; \frac{e^{2}}{k} \; 
\int \!\! d{\bf r}' \; \frac{\theta(R-r')}{|{\bf r -r}'|}
\nonumber \\  &=& \left\{ 
\begin{array}{ll}
4 \frac{e^{2}}{k}
\; n_{0} \; R \; E\left(r^{2}/R^{2} \right), & \mbox{$r < R$} \\
4 \frac{e^{2}}{k}
\; n_{0} \; r \left[ E \left( R^{2}/r^{2} \right) - \left( 1 -
\frac{R^{2}}{r^{2}} \right) \; K\left( R^{2}/r^{2} \right) \right],
& \mbox{$r > R$} 
\end{array} \right.,
\label{vhconst}
\end{eqnarray}
where $K(x)$ and $E(x)$ are the complete elliptic integrals of the first
and second kind respectively. 
The densities associated with the orbitals $\phi_{h}({\bf r})$ 
and $\phi_{N}({\bf r})$ are treated as properly normalized delta functions 
$| \phi_{l}({\bf r}) |^{2} \sim \delta(r - R_{l}) / (2 \pi R_{l}),$
in the terms of Eq.~(\ref{hartreeenergy})
which involve the interaction of any of these two orbitals with the MDD
and  with each other. The self-interaction of the orbitals
$\phi_{l}$'s is treated as that of rings whose electrostatic energy is 
$e^{2}/(2 k C_{l})$ with the capacitance of the ring 
$C_{l} = \pi R_{l}/ [\ln{( 2 R_{l} / \lambda)}]$.  Using 
these approximations in
(\ref{vhconst}) we obtain 
\begin{eqnarray}
\Delta E_{H} &\approx&  \frac{e^{2} 4 N}{k \pi R_{MDD}} \Biggl\{
\int_{0}^{R_{MDD}} \!\! dr' \; 2 \pi r' E\left(r'^{2}/R_{MDD}^{2} \right) 
\left( | \phi_{h}(r')|^{2}  - | \phi_{N}(r')|^{2} \right) \Biggr.
\nonumber \\ 
&+&  \frac{1}{R_{MDD}} \int_{R_{MDD}}^{\infty} \!\! dr' \; 2 \pi r'^{2}  
\Big[ E \left( R_{MDD}^{2}/r'^{2} \right) \Big. \nonumber \\
&-&\Biggl. \Big. \left( 1 - 
\frac{R_{MDD}^{2}}{r'^{2}} \right) \; K \left(R_{MDD}^{2}/r'^{2} \right) 
\Big] \left( | \phi_{h}(r')|^{2} - 
| \phi_{N}(r')|^{2} \right)  \Biggr\} \nonumber \\
&-& \frac{e^{2}}{2 k  C_{h}} - \frac{e^{2}}{2 k  C_{N}}
+ \frac{2 e^{2}}{k \pi R_{N}} K \left(R_{h}^{2}/R_{N}^{2} \right)
\nonumber \\
&=& \left\{ \frac{2 \sqrt{2 N}}{\pi} E\left(\frac{h+1}{N}
\right) - \frac{2 \sqrt{2 (N+1)}}{\pi} 
\left[ E \left( \frac{N}{N+1} \right)  \right. \right. \nonumber \\
&-& \left. \left(1-\frac{N}{N+1}\right)
 K\left( \frac{N}{N+1} \right)\right] 
 -  \frac{1}{2 \pi \sqrt{2(h+1)}} \ln{\left[2 \sqrt{2(h+1)}\right]}
\nonumber \\
&-& \frac{1}{2 \pi \sqrt{2 (N+1)}} 
\ln{\left[2 \sqrt{2 (N+1)}\right]} \nonumber \\
&+& \Bigg.  \frac{2}{\pi \sqrt{2 (N+1)}} 
K\left( \frac{h+1}{N+1} \right) \Bigg\} \frac{e^{2}}{k \lambda}.
\label{dvhapp}
\end{eqnarray}
The variation of exchange-correlation energy between the two states can
be evaluated within a local density approximation 
\begin{eqnarray}
\Delta E_{xc} = \int \!\! d{\bf r} \; \epsilon_{xc}(1) \; \theta(R_{MDD}-r)
\; (n_{MDD}({\bf r}) - n_{r}({\bf r}))&& \nonumber \\
= \int \!\! d{\bf r} \;  \epsilon_{xc}(1) \; \theta(R_{MDD}-r) \;
( | \phi_{h}({\bf r}) |^{2} - | \phi_{N}({\bf r}) |^{2}) 
\approx \epsilon_{xc}(1) = \epsilon_{xc}^{\ast}(1) 
\; \frac{e^{2}}{k \lambda},&&
\label{dexcapp}
\end{eqnarray}
where we have neglected variations in the tail of the MDD and 
we have again replaced the 
orbital densities with $\delta-$functions. $\epsilon_{xc}(1)$ is the
exchange-correlation energy per particle of a uniform electron gas for filling
factor $\nu = 2 \pi n({\bf r}) l^{2} = 1.$ From Eqs. (\ref{lan2mom}),
(\ref{dvhapp}), and (\ref{dexcapp}), we arrive at 
\begin{eqnarray}
E[n_{MDD},\alpha,N] - E[n_{r},\alpha,N] &\approx& \frac{e^{2}}{k \lambda}
\left\{ \alpha (h-N) + \frac{2 \sqrt{2 N}}{\pi} 
 E\left(\frac{h+1}{N}\right) \right. \nonumber \\
&-& \frac{2 \sqrt{2 (N+1)}}{\pi} 
\left[ E \left( \frac{N}{N+1} \right)  \right. \nonumber \\
&-& \left. \left(1-\frac{N}{N+1}\right)
 K\left( \frac{N}{N+1} \right)\right]  \nonumber \\
& - &  \frac{1}{2 \pi \sqrt{2(h+1)}}
\ln{\left(2 \sqrt{2(h+1)}\right)} \nonumber \\
&-& \frac{1}{2 \pi \sqrt{2 (N+1)}} 
\ln{\left(2 \sqrt{2 (N+1)}\right)} \nonumber \\
&+& \left. \frac{2}{\pi \sqrt{2 (N+1)}} 
K\left( \frac{h+1}{N+1} \right) + \epsilon_{xc}^{\ast}(1) \right\}.
\label{deltotapp}
\end{eqnarray}
By equating the left-hand side of Eq.~(\ref{deltotapp}) to zero, 
namely by looking at the transition from the MDD to the reconstructed
edge, one gets the value of $\alpha = \alpha_{1}(N;h)$ 
which characterizes this transition. 
The reconstructed phase occurs when 
$\alpha  <  \alpha_{1}(N; h),$ for a given 
value of $h.$ Therefore 
the position of the first hole in the reconstructed edge is obtained by
{\it maximizing} $ \alpha_{1}(N; h)$ with respect to $h.$ 
This permits also to derive the maximum possible value 
$ \alpha_{1}^{\ast}(N)$ 
for which a transition from the MDD to the reconstructed edge is possible. 
If $N$ is sufficiently large and $h \approx N,$ we have
\begin{equation}
\alpha_{1}(N;h) \approx \frac{1}{N-h} \; \left\{
\frac{2 \sqrt{2 N}}{\pi} \left[ E \left(\frac{h+1}{N}\right) - 1 
\right] + \epsilon_{xc}^{\ast}(1) \right\}.
\label{alphahol}
\end{equation}
Differentiation with respect to $h$ gives 
\begin{eqnarray}
\frac{\partial  \alpha_{1}}{\partial h} &\approx&
\frac{1}{N-h} \left\{ \frac{2 \sqrt{2 N}}{\pi (N-h)} \left[ 
E\left(\frac{h+1}{N}\right) - 1 \right] \right. \nonumber \\
&+& \Biggl. \frac{\epsilon_{xc}^{\ast}(1)}{N-h}  
- \frac{1}{2 \sqrt{2 N}} \;
_{2}F_{1}\left(\frac{3}{2},\frac{1}{2};2;\frac{h+1}{N}\right) \Biggr\}
\label{dadhapp2}
\end{eqnarray}
with $_{2}F_{1}(a,b;c;z)$ the hypergeometric function.
Under the hypothesis --- which we shall show is valid --- of $h \approx N,$
the ratio $(h+1)/N$ is a number smaller than one, but close to unity. 
Therefore --- in order to find  
an expression for the change of angular momentum associated with
the reconstruction of the edge --- it makes sense
to use an expansion of the hypergeometric function in terms 
of $1-z,$ and retain the lowest order terms.
From~\cite{abrmsteg70}
\begin{equation}
E[x] \approx 1 + (1-x) [a_{1} - b_{1} \ln{(1-x)}]; a_{1}=0.44325141463,
\; b_{1} = 0.24998368310,
\label{ellipteappr}
\end{equation}
and recalling the representation~\cite{erdel53}
\begin{eqnarray}
_{2}F_{1}[a,b;a+b+l;z] &=& \frac{\Gamma(l) \Gamma(a+b+l)}{\Gamma(a+l) 
\Gamma(b+l)} \; \sum_{n=0}^{l-1} \; \frac{(a)_{n} \; (b)_{n}}{(1-l)_{n}
n!} \; (1-z)^{n} \nonumber \\
&+& (1-z)^{l} (-1)^{l} \; \frac{\Gamma(a+b+l)}{\Gamma(a) \; \Gamma(b)}
\nonumber \\
&\times& \sum_{n=0}^{\infty} \; \frac{(a+l)_{n} \; (b+l)_{n}}{n! \; 
(n+l)!} \; [k_{n} - \ln(1-z)] \; (1-z)^{n}
\label{hypgrep}
\end{eqnarray}
for the hypergeometric function, valid whenever 
$c=a+b+l,l=0,1,2,\ldots,$ 
we obtain --- retaining only the lowest order term in $(1-z),$
\begin{eqnarray}
\frac{\partial \alpha_{1}}{\partial h} &\approx&
\frac{1}{(N-h)^{2}} \left\{ \frac{2 \sqrt{2}}{\pi} \frac{N-h}{\sqrt{N}} \;
\left[ a_{1}  + b_{1} \; \ln{\left(\frac{N}{N-h}\right)}\right] 
 + \epsilon_{xc}^{\ast}(1) \right\} \nonumber \\
&-& \frac{1}{N-h} \; \frac{\sqrt{2}}{\pi \sqrt{N}} \; \left[ 2 \ln{2}-1
+ \frac{1}{2} \ln{\left(\frac{N}{N-h}\right)}\right].
\label{dalphadafin}
\end{eqnarray}
In Eq.~(\ref{hypgrep}) 
\begin{equation}
k_{n} = \psi(n+1) + \psi(n+1+l) - \psi(a+n+l) - \psi(b+n+l).
\label{kndef}
\end{equation}
$\psi$ is the logarithmic derivative of the gamma function, while
$(a)_{n}= \Gamma(a+n)/ \Gamma(a),$ is Pochhammer's symbol.
Application of the condition $\partial \alpha_{1}/ \partial h = 0,$
to Eq.~(\ref{dalphadafin}) results in 
\begin{equation}
N-h \approx - \frac{\pi \epsilon_{xc}^{\ast}(1)}{2 \sqrt{2} \left[ a_{1} - 
\frac{1}{2} (2 \ln{2}-1)\right]} \; \sqrt{N},
\label{nminh}
\end{equation}
where we have used $4 \; b_{1} \approx 1.$
The value  of  $\epsilon_{xc}(1) = -0.7015 e^{2}/(k l),$ gives
$N-h \approx 3.115 \; \sqrt{N}.$ The same scaling behavior  was found
numerically by Oaknin {\em et al.,}~\cite{oaknin95} with the only
difference in the value of the prefactor, being 2 in their case. The
difference between our and their results might be a consequence of the
fact that the latter have been obtained for systems smaller than those
in which the  asymptotic behavior of Eq.~(\ref{nminh}) applies. 
We also notice that introducing the result of Eq.~(\ref{nminh}) in 
Eq.~(\ref{alphahol}), the latter 
agrees with the behavior for $\alpha_{1}^{\ast}$ predicted by
Chamon and Wen,~\cite{chamon94} except for the value of the numerical
factors. The difference in the prefactors is due to the fact that while
in our case we obtain it from an asymptotic behavior, Chamon and Wen 
determined it in terms of a fitting procedure.
In Fig. \ref{holepredfig} we present the increase of angular momentum with
respect to the MDD after the edge reconstruction takes place,
scaled by the square root of the number of electrons, as a function of
the number of electrons in the dot, obtained by numerically finding the 
maximum with respect to $h$ of $\alpha_{1}(N; h)$ as obtained by
equating to zero the left-hand side of 
Eq.~(\ref{deltotapp}). The asymptotic value
of Eq.~(\ref{nminh}) for large dots --- represented by a dashed line in
Fig. \ref{holepredfig} --- is approached very slowly, basically for
values much larger than those considered in the figure.

The other boundary of the MDD region of the phase diagram is derived in
an analogous fashion. Here the new phase is described by flipping the
spin of one electron in the MDD and putting it into a state with angular
momentum $0 \leq l \leq N-2.$ 
The transition between the two phases occurs when
\begin{eqnarray}
\alpha = \alpha_{2}(N; l) &\approx& - \frac{1}{N-(l+1)} \; 
\left\{ \frac{2 \sqrt{2 N}}{\pi}
\left[ 1 - E\left(\frac{l+1}{N}\right) \right] \right. \nonumber \\
&-& \frac{\ln{ \left(2 \sqrt{2 N} \right)}}{2 \pi \sqrt{2 N}} -
\frac{ \ln{ \left(2 ( \sqrt{2 (l+1)})\right)}}{2 \pi \sqrt{2 (l+1)}} 
\nonumber \\
&+& \left. \frac{2}{\pi \sqrt{2 N}} \; K\left(\frac{l+1}{N}\right) 
+ \epsilon_{xc}(1)  \right\}.
\label{alphmin}
\end{eqnarray}
In Eq.~(\ref{alphmin}), the last term accounts for the change in
exchange-correlation energy, while the remaining ones are purely
electrostatic, and represent the energy required for moving a
charge distribution located around $R_{N-1}$ to a neighborhood 
of $R_{l}.$
We see that this region of the phase diagram is dominated by
electrostatics, exchange-correlation effect being of higher order in
$1/\sqrt{N}.$
The expression for Eq.~(\ref{alphmin}) has to be minimized
with respect to $l,$ in order to obtain the first configuration
that gives the ground state when the MDD is lost for increase of 
$\alpha.$ Considering the terms with leading order in 
$N,$ we arrive at
\begin{eqnarray}
\alpha_{2}(N; l) &\approx& - \frac{1}{N-(l+1)} \; \left\{ 
\frac{2 \sqrt{2 N}}{\pi} \left[ 1 - E\left(\frac{l+1}{N} \right) \right]
+ \epsilon_{xc}^{\ast}(1) \right\} \nonumber \\
&\approx& - \frac{1}{N-(l+1)} \left\{ - \frac{2\sqrt{2 N}}{\pi}
\; \frac{N-(l+1)}{N} \;  \left[ a_{1} + b_{1} \; 
\ln{\left(\frac{N}{N-l-1}\right)} \right]
+ \epsilon_{xc}^{\ast}(1) \right\}.
\label{alphminapp}
\end{eqnarray}
Eq.~(\ref{alphminapp}) is monotonically increasing with $l.$ 
It follows then that  the angular momentum that 
minimizes Eq.~(\ref{alphminapp}) is  $l=0.$ Therefore
\begin{eqnarray}
\alpha_{2}^{\ast}(N) &\approx& - \frac{1}{N-1} 
\left\{ \frac{2\sqrt{2N}}{\pi}
\left[1 - E(1/N) \right] + \epsilon_{xc}^{\ast}(1) \right\} \nonumber \\
&\approx& \sqrt{2} \; \frac{\pi-2}{\pi \sqrt{N}} - 
\frac{\epsilon_{xc}^{\ast}(1)}{N}.
\label{alphaminl0}
\end{eqnarray}
From the previous analysis we conclude that the MDD is a ground state
whenever 
$\alpha_{1}^{\ast} < \alpha < \alpha_{2}^{\ast}.$ 
Since the expressions for $\alpha_{1}^{\ast}$ and 
$\alpha_{2}^{\ast}$
depend {\em only} on the number of electrons $N,$ the maximum value
$N_{c}$ for which the MDD is a ground state for some values of $\alpha$
is obtained from 
$\alpha_{2}^{\ast}(N_{c})=  \alpha_{1}^{\ast}(N_{c}).$
By solving the previous equation by means of (\ref{alphahol}) and
(\ref{alphaminl0}), we get $N_{c} = 222.$ If instead we use the
expressions for the $\alpha^{\ast}$'s as obtained by equating to zero
(\ref{deltotapp}) and from (\ref{alphmin}), finding then their maximum
with respect to $h,$ and minimum with respect to $l$ respectively, we 
obtain $N_{c} = 160.$

\begin{figure}
\centerline{\psfig{figure=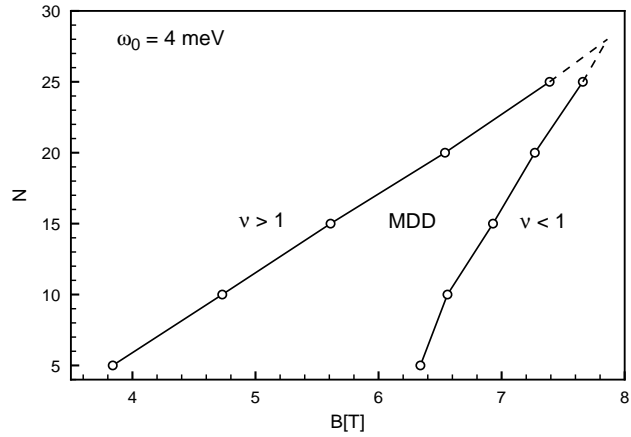,width=3.5in}}
\caption{{\it B-N phase diagram of the MDD for $\omega_{0} = 4 {\rm
eV}$  obtained  solving the Kohn-Sham equations as in 
Ref.~\protect\onlinecite{ferconi94}.}}
\label{mddphdiag}
\end{figure}

The above discussion was limited to the lowest Landau level. 
We now turn to considering the effects coming from the inclusion of higher
Landau levels.  
In this case the full Kohn-Sham equations for the quantum
dot must be solved.~\cite{ferconi94} The results for the B-N
phase diagram of the MDD are presented in Fig. \ref{mddphdiag}.
The most prominent feature  (in comparison to the LLL approximation)
is a narrowing of the window of magnetic fields for which the MDD
is the ground state. In particular, $N_{c} \approx 28,$ when 
$\omega_{0} = 4 {\rm meV.}$

\begin{figure}
\centerline{\psfig{figure=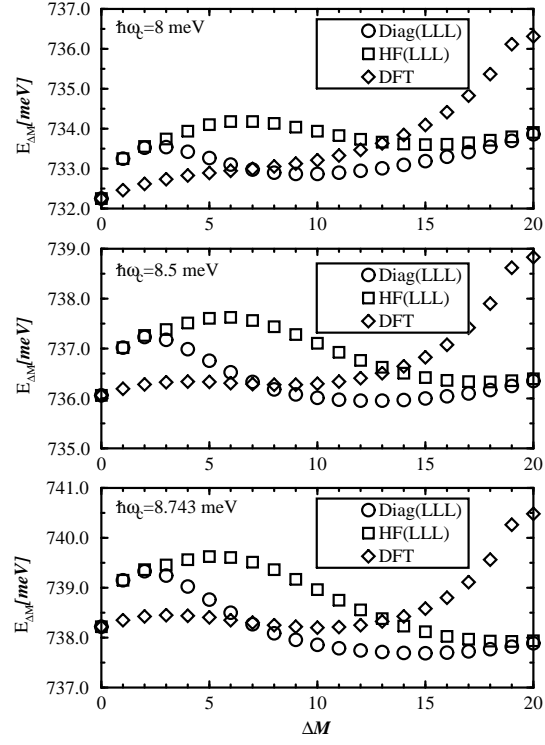,width=3.5in}}
\caption{{\it Evolution of the energy for the MDD and the ``one-hole''
states as a function of the magnetic field for a 20 electron quantum
dot. $\omega_{0} = 3 meV.$}}
\label{cor_effectrec}
\end{figure}

Therefore we see that Landau level mixing is essential to an accurate 
determination of  the  correct values at which the transitions take place. 
Correlation effects are manifested by giving a more compact
quantum dot with respect to what is given by Hartree-Fock theory,
which includes only exchange. Moreover,
the values of the angular momentum of the reconstructed edge as
predicted by DFT are in better agreement than HF with exact
diagonalization  calculations. This is represented in
Fig. \ref{cor_effectrec}, where we give the values of the energy of
a MDD with a hole in it (i.e., a MDD on the verge of reconstruction), as a
function of the position of the hole, for three different values 
of the magnetic fields,  as evaluated by numerical diagonalization
within the LLL, by DFT,  and 
by HF.  $\Delta M$ is the increase of angular momentum for ``one-hole''
states with respect to the angular momentum of the MDD.
The fact that DFT departs from the exact diagonalization results for
large $\Delta M$  is not surprising, since those states correspond to
excited states, where DFT is not applicable.
Of course these states do not affect the B-N phase diagram for the MDD.

In conclusion, we have shown that correlation effects which are incorporated in
DFT give rise to a more compact
reconstructed edge than the one predicted by HF theory.
These results are in better agreement with those of exact diagonalization
studies.  A simplified model reproduces the main physical 
traits involved in the
phase diagram of the MDD, which is determined, within the model, by means
of analytical expressions.

M.\ F.\ acknowledges financial support from ONR  under grant
No. 00014-96-1-1042. G.\ V.\ acknowledges financial support from NSF
under grant No. DMR-9403908.

\widetext
\end{document}